\documentclass[10pt,letterpaper]{article}
\usepackage{iccv}
\usepackage{times}
\usepackage{epsfig}
\usepackage{graphicx}
\usepackage{amsmath}
\usepackage{amssymb}
\usepackage{booktabs}
\usepackage{longtable}
\usepackage{caption}
\usepackage[utf8]{inputenc}
\usepackage[breaklinks=true,bookmarks=false]{hyperref}

\iccvfinalcopy
\title{Holistic Evaluations of Topic Models}
\author{
Thomas Compton\\
University of York\\
York, United Kingdom\\
{\tt\small thomas.compton@york.ac.uk}
}

\begin{document}
\maketitle

\textbf{Abstract}

Topic models are drawing increased commercial and academic interest for their ability to summarise large quantities of unstructured data (Hosseiny Marani and Baumer, 2023; Egger and Yu, 2022; Li et al., 2023; Madrid-García et al., 2024). As unsupervised machine learning approaches, they offer a method of studying data for researchers and understanding the important points of large amounts of text to general users. At the same time, they risk becoming a ‘black box’ (Wang et al., 2024), where researchers add their data and uncritically assume the output results are a ‘correct’ summarisation of the data. In this article, I wish to approach the evaluation of topic models from a database perspective, drawing inferences from outputs from 1140 runs of BERTopic models. The goal will be to explore potential trade-offs in optimising certain aspects of the topic model and to consider what these findings mean for how we should interpret topic models.
\vspace{1em}

\section*{1 Topic Models}

First, it is important to clarify the architecture of topic models. With BERTopic, for example, the approach is stochastic (Kumar, Karamchandani and Singh, 2024), relying on UMAP and HDBScan (Gokcimen and Das, 2024; Grootendorst, 2022). UMAP is used to reduce the embeddings, which can be produced by sentence transformers. Then HDBScan clusters the vectors into meaningfully distinct groups (Lin et al., 2024). For research, I shall focus on UMAP and HDBScan because they are the sources of variation in the runs. In contrast, the use of Count Vectorisation to extract ngrams is deterministic, meaning it will produce the same top 10 ngrams for the same cluster.

It is important to clarify BERTopic’s advantage against a deterministic model such as K means. If one were to use K means clustering, the variation would come from the user inputting the K value, which is the number of clusters pre-set. After this has been set, it should be the case that re-running the model produces the same results. However, the trade-off is that K means does not necessarily provide meaningful topics. So, HDBScan is sacrificing repeatability for better quality clusters. The question emerges from this: how much of an issue is repeatability, and how improved is HDBScan compared to K means?

BERTopic has the advantage of being user-controllable via parameters (Liu, 2024). Table~\ref{tab:error-size} illustrates the relationship between `min\_cluster\_size` (a parameter in HDBScan used to tune the clustering size), `min\_topic\_size` (used in BERTopic to limit topic minimums), and `n\_neighbors` (used in UMAP to define local structure). If this were a deterministic model, the table would help determine the best parameters. Yet because the model is stochastic, re-running it with the same parameters may yield different results. The `error\_size` column shows how many sentences were assigned to the `-1` topic — a signal that no meaningful topic could be found.

\begin{table}[htbp]
\centering
\small
\caption{Overview of Relationship between Parameters and Error Size}
\label{tab:error-size}
\begin{tabular}{cccc}
\toprule
\textbf{min\_cluster\_size} & \textbf{min\_topic\_size} & \textbf{n\_neighbors} & \textbf{error\_size} \\
\midrule
10 & 30 & 5 & 6829 \\
20 & 35 & 5 & 7275 \\
10 & 50 & 5 & 7326 \\
25 & 35 & 5 & 7353 \\
\bottomrule
\end{tabular}
\end{table}

\section*{2 Existing Studies on Topic Model Evaluation}

Given the popularity of topic models and range of potential options it is unsurprising to find many comparisons between topic models such as BERtopic, LDA, and/or Top2Vec (An, Oh and Lee, 2023; Egger and Yu, 2022; Li et al., 2025).

\begin{table}[htbp]
\centering
\small
\caption{Evaluation Terms in Literature}
\label{tab:eval-terms}
\begin{tabular}{lcc}
\toprule
Term & Frequency & Percentage \\
\midrule
word embedding & 33 & 61 \\
Pointwise Mutual Information & 11 & 20 \\
cosine similarity & 15 & 28 \\
qualitative & 23 & 43 \\
\bottomrule
\end{tabular}
\end{table}

This table suggests that BERTopic users are using different methods to evaluate their model with no clear most prominent metric in this sample. PUV (Pairwise Uniqueness Value) did not appear in the corpus. This metric evaluates the overlap in ngrams between topics. If topics share many ngrams, it suggests each topic is not distinct and therefore the clustering was not effective in semantic differentiation. PMI, which can be normalised, compares the probabilities of the ngrams co-occurring in the corpus against each ngrams probability of occurring in the corpus (Ranaweera et al., 2025). Therefore, this is not comparing topics, but an attempt to gauge how effectively the topic model has grouped together co-occurring terms. However, BERTopic uses C-TF-IDF, which means that it is not outputting the most common ngrams from each cluster by attempting to find contextually significant terms. This means that this co-occurrence may suffer as BERTopic may be choosing low-frequency terms as an internal buffer against topic overlap, especially in smaller corpora.

Gan et al., (2024) uses cosine similarity to compare the semantic differences between topics (WE). This can be done through converting the ngrams output for each topic into a string then embedding it with sentence transformers. After this the distance between the vectors of each pseudo-sentence can be compared. Alternatively, each ngram could be converted into a vector and the usual PUV approach can be used with the advantage of increased sensitivity to semantic similarity. This method is computationally intensive, as compared to PUV. The issue is that each model uses a different approach to extract ngrams, so the issue becomes: are we testing these packages as wrappers or should the focus be normalising the ngram selection process to ensure that the clustering process is being tested.

Equally, topic coherence values can be computed using a similar approach but comparing ngrams within topics using cosine similarity (Khodeir and Elghannam, 2025). As with PMI and PUV, the focus is on the ngrams which is the output from the models many users may highlight. At the same time, using UMAP projections, users are able to inspect the topic overlaps through looking at the clusters. This approach would allow for inspection of how effective the clustering process has been, instead of focusing on evaluating the ngram selection. This is valuable because the ngram outputs provide useful information about the topics, but for researchers interested in understanding limitations to the clustering approach, they do not provide much useful information about how effectively the models have been clustering. In this corpora, there was no mention of using Gini Coefficients, for example, to test the distributions of clusters.

\begin{table}[htbp]
\centering
\small
\caption{Topic Model Literature}
\label{tab:literature-terms}
\begin{tabular}{c p{5cm} c c}
\toprule
Rank & Term & Frequency & Percentage \\
\midrule
0 & Latent Dirichlet Allocation & 47 & 87\% \\
7 & UMAP & 43 & 80\% \\
2 & HDBSCAN & 40 & 74\% \\
1 & top2vec & 32 & 59\% \\
3 & k-means & 15 & 28\% \\
5 & min\_cluster\_size & 5 & 9\% \\
6 & n\_neighbors & 5 & 9\% \\
4 & min\_topic\_size & 4 & 7\% \\
\bottomrule
\end{tabular}
\end{table}

\clearpage

Looking at terms associated with topic modelling, it is clear the corpora contains many articles comparing BERTopic to LDA or top2vec. These comparisons could be tests to see which performs better, or they could be triangulations where researchers attempt to gain more information about their corpora by using both. The use of comparative approaches is not surprising as the evaluative metrics seem to focus on testing the outputs of models so can conveniently move between models. Although, this will be limited by how each model uses a different approach at extracting ngrams. Moreover, this approach is a good step because there are no ground truth topic labels, so comparing outputs can increase confidence in findings. On the other hand, if each model output is not optimised, the triangulation may be giving users a false sense of confidence in their findings.

To explain this tension further, I shall look at how BERTopic specifically is being used. While UMAP and HDBScan are being identified as the steps for topic modelling, there is less discussion of the key metrics for fine-tuning the outputs. As we demonstrated in the last section, the changing of these metrics can have a significant effect on the error size. This means that these topic models are theoretically ‘plug and play’, but in practice, outputs from one dataset could vary in quality. However, to test this, it will be important to establish by what standard we are measuring the denigration of quality. This article will not assume ‘plug and play’ usage is inherently poor, instead seeking to understand the potential limitations of both manual interference and allowing the model to run without customisation.

Therefore, the key results for this initial review is that metrics for evaluating topic models are not always guaranteed in studies. There is a focus on studying the outputs of the model, not necessarily the clustering process. This is useful for comparing between topic model approaches. However, the lack of discussion of fine tuning parameters suggests these comparisons are mainly being doing in a ‘plug and play’ format. However, it is not yet possible to argue this is a limitation. From this, I argue there is space to consider how far changing the parameters of BERTopic has an effect on the outputs, what level of trade-off might be expected. Also, there could be scope for consideration of new metrics which are sensitive to different aspects of the topic modelling pipeline.

\section*{3 Developing Evaluation Metrics}

Before discussing the metrics I propose for evaluating topic models, it will be important to establish the features of an ideal topic model. On one hand, this is a difficult task because text corpora are non-normalised and highly varied data. Therefore, the expectations cannot be too fixed and risk losing sensitivity to variation. On the other hand, there are metrics based on the model outputs that can resolve this issue. Since the model is already normalising the data and providing a quantitative output, it is possible to focus on the outputs of the model and make comparisons between different models.

First, a topic model should not have skewed topic counts. While this could be indicative of skewing of topic distribution within corpora, it is plausible topic models with poor distributions are instances of models improperly clustering sentences. It could be that this correlates with decreasing the -1 number of sentences, as I shall explore in the next section. Therefore, a model that has a high gini coefficient is likely to have a less than ideal distribution of topic. Thus, it will be less readable, especially as a UMAP projection and potentially less valuable because the topic model has improperly assigned sentences to clusters, likely to increase their size.

\textbf{Gini Coefficient Formula}
\[
G = 1 - \sum_{i=1}^{n} p_i^2
\]
Where $n$ is the total number of topics; $p_i$ is the proportion of sentences assigned to topic $i$. This version of the Gini coefficient is widely used to measure inequality in distribution, here applied to sentence-topic assignments. A higher Gini value means more imbalance, indicating that some topics dominate while others are underrepresented.

Next is keyword or ngram value. It is important that topic models contain a reasonable quantity of high frequency ngrams. This is a method for testing how effectively the topic model summarises the data. This is a novel approach I have developed, totalling the frequencies for each ngram produced by the model for the top 20 topics. Again, there cannot be an absolute metric, but this value can be compared with other metrics and can be compared with other models more broadly. Both the gini coefficient and keyword value are not definitive metrics, insofar as it would be difficult to guarantee a threshold for securing a good model. However, they do provide a broader picture for understanding variation within topic models and gauging how far each topic model fits certain criteria.

Let:

- $T_{\text{all ngrams}}$ be the total number of ngrams in the corpora, calculated using \texttt{scikit-learn} \texttt{CountVectorizer}
  
- $s(w_i)$ be the c-TF-IDF score of the $i$-th ngram ($w_i$), as assigned by \texttt{BERTopic}

Then the Ngram Frequency Score (NFS) is defined as:
\[
\text{NFS} = \frac{1}{T} \sum_{i=1}^{T} s(w_i)
\]

Another metric is a fuzzy topic stability checker. This metric will require previous runs to create a lookup table of previous topics. Then the top 20 topics can be compared by the frequencies in the table, based on WER (word error rate), to give a fuzzy matching approach which allows for total scores based on similar topic names. This approach is similarly only indicative and comparative. Yet, it does provide the opportunity to explore the issues relating to topic stability. It is based on how BERTopic names topics using the top ngrams. Another approach could compare all the ngrams or the model. This would have the advantage of being a more extensive search. At the same time, it would not account for the values of each ngram, treating them all equally. Either approach would be limited, as would any topic stability in the vagueness of what exactly the topic is. Whether a topic is the name output, or whether it is best described by its top 10 words.

PUV appears not to feature in the corpus of literature. However, attempts to compare the ngrams are present. This is why I will build on their contributions to offer an Ngram Uniqueness Value (NUV) as a metric to determine how many lemmatised ngrams are occurring multiple times within a topic model. This approach is lightweight compared to a semantic comparison because it does not rely on embedding each word. This makes it ideal for testing over multiple runs, even if this approach will not be comparing as rich data as WE comparisons.

Let:

- $T$ be the total number of n-grams across all selected topics (e.g., top 10 n-grams $\times$ 20 topics, so $T = 200$)
  
- $N$ be the number of non-unique n-grams, i.e., n-grams that appear more than once across different topics

Then, the Non-Unique Value (NUV) is defined as:
\[
\text{NUV} = \frac{N}{T}
\]

\begin{table}[htbp]
\centering
\small
\caption{Comparing BERTopic, Top2Vec and LDA}
\label{tab:comparison}
\begin{tabular}{lccc}
\toprule
Metric & BERTopic & Top2Vec & LDA \\
\midrule
PUV & 0.968 & 0.926 & 0.947 \\
Coherence (c\_npmi) & -0.075 & -0.270 & -0.062 \\
\% Appearance & 9.2\% & 15.2\% & 100.0\% \\
Gini coefficient & 0.139 & 0.114 & 0.140 \\
\bottomrule
\end{tabular}
\end{table}

This initial comparison explores BERTopic and Top2Vec using ‘all-MiniLM-L6-v2’, the model which is the default embedding model for bertopic (Gokcimen and Das, 2024).  It is referred to in 17\% of the literature corpus. Therefore, the goal of this table is to give a baseline of what a user might receive in quality from each model, comparing the top 20 topics from each. To standardise this, the LDA, using Genism’s model, had 20 topics selected. This model compares the top 20 topics from each model, with \% appearance referring to the coverage of the model as a percentage of the total sentences. BERTopic scores best in PUV and NUV, demonstrating the model produces distinct ngrams between topics (PUV) and distinct ngrams overall. LDA performs best at Topic 20 Size, Appearance, Coherence and KFS. This means that LDA’s topic clusters are larger, which is unsurprising given the high coverage. Yet, this is not providing a trade off in coherence. Although the much increased KFS could indicate the ngram outputs are being diluted because of the cluster size. Top2Vec has the best Gini, which indicates the most even distribution of topics, but each value is notably low.

From these metrics, it might be fair to argue BERTopic, tentatively, outperforms Top2Vec by providing more coherent but smaller clusters. On the other hand, these preliminary metrics do not provide a broader picture of the scope and potential damage this trade-off may cause. What has made BERTopic the model for this study, over Top2Vec, also involves the modularity of the wrapper, as it is designed to facilitate. Compared to Grootendorst (2021), each of these coherence score underperforms his tests. Although, his test of BERTopic uses ‘all-mpnet-base-v2’ a model with benchmarks superior to ‘all-miniLM-L6-v2’. Medvecki et al. (2024, p.6) Found -.058 using the similar ‘paraphrase-multilingual-mpnet-base-v2’, demonstrating a better score than -0.075 for our test.

These proposed metrics will be explored in the next section to see how they interact with each other and how far they are useful in creating more valuable topic models. As already demonstrated in the previous section, repeating topic models using randomised parameters can lead to optimised metrics at the expense of user time. To ensure continuity, a similar approach will be taken.

\section*{4 Descriptives}

\begin{table}[htbp]
\centering
\caption{Large Corpus RUN Descriptives}
\label{tab:large_corpus}
\begin{tabular}{lrrrr}
\toprule
\textbf{Index} & \textbf{Mean} & \textbf{Standard Deviation} & \textbf{Min} & \textbf{Max} \\
\midrule
Error\_Size          & 56281.51 & 4604.76  & 41846.0 & 62180.0 \\
Keyword\_Freq\_Score &     0.18 &    0.01  &     0.15 &     0.20 \\
Gini\_Score          &     0.46 &    0.10  &     0.32 &     0.76 \\
Topic\_20\_Size      &   793.05 &  251.98  &   117.0  &   1421.0 \\
\bottomrule
\end{tabular}
\end{table}

\begin{table}[htbp]
\centering
\caption{Total BERTopic Runs}
\label{tab:runs_summary}
\begin{tabular}{lrrr}
\toprule
\textbf{Run Name} & \textbf{Total Values} & \textbf{Unique Values} & \textbf{Unique Percentage} \\
\midrule
PUV (large)   & 342 & 55  & 16\%  \\
RUN (large)   & 301 & 41  & 14\%  \\
PUV (small)   & 405 & 113 & 28\%  \\
RUN (small)   & 92  & 92  & 100\% \\
\midrule
\textbf{Total} & \textbf{1140} & \textbf{301} & \textbf{26\%} \\
\bottomrule
\end{tabular}
\end{table}

From these metrics, it is clear Gini Coefficient, the size of the 20th topic and error size saw more variation than keyword frequency. Gini and topic 20 size should have a relationship, which will be explored in the next section. Keyword frequencies suggest most models will have reasonable consistency on this metric. However, the Gini coefficient indicates there can be issues with the sentence distribution. Therefore, these data suggest there can be variation between models a user may want to avoid. At the same time, the lack of unique values from this number of runs indicates the parameter range was able to keep the output reasonably consistent.

Therefore, there will be diminishing returns for users who run more models. The issue for research is that the parameters interact such that even an approach that only used one parameter combination once would still see repeated values.

\section*{5 BERTopic Statistics}

Before discussing the findings, it is worth discussing the context of this data. Knowing that BERTopic is stochastic because of UMAP and HDBScan (Kumar, Karamchandani and Singh, 2024), these tests will be limited by being a partial collection of results. It may be more appropriate to say, it is difficult to gain a full set of possible outputs because of the stochastic nature and variation between datasets. To resolve this issue, I have used 2 datasets of different sizes. This will provide the opportunity to discuss differences between them and possible reasons. The other limitation in this data is the nature of textual corpora. Each corpora is substantially different, especially regarding how far high frequency terms are used within thematically consistent sentences. Essentially, one might expect some corpora to score higher on ngram values because of the content of the sentences, not the BERTopic model. Grootendorst (2022) also found varying outputs from BERTopic between corpora. This means between corpora comparisons should be done with caution, and the focus should be on metrics which should remain reasonably consistent across corpora. Therefore, there are limitations to this statistical approach. However, it should be useful to give guidance to users of BERTopic surrounding general issues, which should not be interpreted as inherent flaws with BERTopic.

Beginning with the Gini Coefficients, this metric is simple to compare across corpora because it is evaluating the model outputs. The small corpora found Spearman’s Rho -0.55, $p = 0.0000$. The large corpora found Rho 0.57, $p = 0.0001$. Both results demonstrate a statistically significant relationship between these metrics, indicating that with more sentences in the -1 topic, greater imbalances occur within the number of sentences in each topic. It is reasonable to argue, therefore, that models which have had their coverage of sentences increased by changing the parameters lead to more sentences falling into the top topics at the expense of the latter topics.

This is substantiated by the relationship of the size of the 20th topic and the error size. With the smaller corpus, this relationship is less defined. Whereas, the relationship is more significant in the larger corpus. This may be related to there being more variation of the 20th topic size in the larger model. With more sentences available to be clustered, there may lead to more significant issues of skewness in larger corpora. On the other hand, the most prominent finding is that it appears there is evidence the primary effect of decreasing the number of sentences in the -1 topic is increasing the number of sentences in the top topics. Therefore, this suggests the quality of the top topics may be decreased by this manual increasing of the sentences covered by the model.

The results of the ngram frequency value differed between corpora. With the large corpora, there appears to be evidence of a curve. As the error size increases, at first, the ngram value increases but stops increasing around the mean. Whereas, the smaller corpora saw a more consistent increase, with a weak correlation. This suggests that there may be a relationship between these metrics, but it may be more dependent on the corpus. This issue was previously discussed, where it may be the case that this metric can be useful for gauging the representation of the model within a corpus, but is not useful across corpora. This would make it difficult to judge what a reasonable score should be for this metric. Instead, it may be more appropriate to consider it more of an indicator of a poor model if the output is below the mean.

With the NUV value, there appears to be evidence of repeated ngrams entering the models as the -1 topic is increased. This relationship is counter-intuitive as one might expect the opposite relationship to be the case as increasing the error size should lead to worse clustering, with higher frequency ngrams outweighing the more topic coherent ngrams. In the same corpus the relationship was Spearman Rho = -0.22, $p = 0.0195$. There appears to be a significant relationship at the 0.05 threshold but not at 0.01. However, the negative correlation is quite weak. The change from 0.78 to 0.74 with 200 ngrams from the top 20 topics, this would indicate a rough change of 8 words. So it is not likely to be a noticeable change for users. So, in this metric, there is little evidence of significant worsening of model quality from NUV. On the other hand, NUV is quite a limited metric in only comparing where there are overlapping ngrams between topics. It does not consider semantic overlap between topics, so there may be a negative effect not considered.

With the larger corpora, the results were Spearman Rho = 0.18, $p = 0.1965$. This indicates a reasonably consistent result, without significance at 0.05. Therefore, there is significance in one dataset but not the other, which suggests there may not be a significant relationship. The mean for the NUV is 0.81 with a standard deviation of 0.01 to 2dp. This indicates quite a consistent output. The small corpora has a mean of 0.76, and standard deviation of 0.01. The score could be higher for the larger corpora because of the larger amount of sentences leading to more distinct clusters. On the other hand, textual data is too inconsistent to make meaningful comparisons based on the presently discussed data.

The metrics indicate a clear trade-off when maximising the model coverage for gini coefficients. On the other two metrics, the relationships were not as meaningful, with there being less reason to argue model coverage would lead to a difference in model quality. Therefore, a user could argue they are willing to sacrifice a more even distribution of sentences for larger coverage. This argument would be reasonable based on the evidence and the goals of their research. Especially, if a user is interested in using a UMAP projection, this will be an issue for their approach as the skewness could make this representation less readable. The NUV scores indicate that this is not a significant source of issue within BERTopic models. Whereas, the ngram frequencies saw more variation. In this case, equally, it will be a decision for a user to decide whether this metric is worthy of increasing.

\section*{6 Discussion}

While the last section could conclusively discuss trade-offs, the question of the validity of each of these metrics may still be dubious. This is because topic skew, for example, may create some issues with improper categorisation. However, we are measuring the effect of the problem, not the problem itself. To do so would be challenging because there is no ground truth. There is no correct set of topic labels and correct identification of which topic each sentence should be clustered into. So these metrics can provide comparative insight on certain features between models. Therefore, there requires more debate on how these metrics can be used and consider the advantages and disadvantages of these use cases.

From the broader perspective, using a topic model and using qualitative judgement, perhaps alongside topic coherence and diversity. They may compare these outputs to other models. For example, Grootendorst (2022) compares multiple runs of BERTopic and compares it to other LDA and Top2Vec (see also Kumar, Karamchandani and Singh, 2024), but does not focus on fine-tuning the parameters. As such, the assumption is that repeated runs insulate against poor quality models caused by the model being unable to appropriately dictate the correct parameters. Of course, this is already part of the difficulty because there is no guarantee that one model will be better or worse than others, especially when relying on default settings. Through reducing the range of randomly chosen variables, changes can be made to the metrics established. However, one first has to believe them to be credible. Therefore, there is no shorthand solution for what a good model is without multiple runs to compare. At the same time, encouraging users to compare multiple runs increases the computational demand and may deter potential users. If they are using the topic model for its convenience in summarising a dataset, they will be disappointed to be delayed.

Worse yet, metrics such as error size and topic 20 size will be corpus dependent. This means that producing a rough guide will be inappropriate. Whereas sentence appearance, gini coefficient and keyword frequency are all standardised. On the other hand, creating a list of ideal frequencies could be useful for a guide but cannot be treated as hard rules because of variation between corpora.

\section*{7 Limitations}

If part of the gambit of this article was exploring whether it is possible to statistically evaluate BERTopic models, it would be fair to state the results are mixed. The relationship between the number of sentences in the -1 topic and gini coefficient was clear. However, the relationship between this error size and the NUV score or keyword frequencies was inconclusive. Unfortunately, more runs would not improve this issue. Instead, the problem is that repeated runs produce similar results. Where this approach suffers is that the error size essentially serves as an index for each model with the error size dictating the results of the metrics consistently. Essentially, a model that has the same error size is the same model. This is because parameters do not dictate the error size. Therefore, the only way to find the error size is to run the model. So, it would be impossible to pre-establish the runtime for an approach of this sort and to approach an exhaustive list of possible outputs.

Moreover, there is a danger of creating a dataset of unusable models to test BERTopic. This would lead to an analysis on models users would most likely reject. Therefore, it might be evaluating BERTopic as a package, but not evaluating possible models produced by users. That is, we would be creating models for the sake of a dataset instead of considering what sort of variation a user might experience in their usage of BERTopic. This is the advantage of the approach taken in this article. We have used possible models that a user might reasonably consider for their project. We have not analysed models a user might reject, with our approach not storing outputs from models with less than 20 topics.

The evaluation of BERTopic models cannot lose sight of the purpose of BERTopic models. Moreover, these metrics are designed to provide insights but to not dictate what a ‘good’ model is. This decision will be, ultimately, qualitative. Furthermore, any attempt to evaluate BERTopic models will be time-consuming and inefficient. We undertook 90 runs with NUV tests and produced 10 unique NUV values. For most users, an approach which includes 5 runs and chooses the best may be what reduces their inconvenience. At the same time, it is plausible 5 runs could include similar values. Ultimately, this is the inherent flaw with using a stochastic model (Kumar, Karamchandani and Singh, 2024), that the chance will always be that the comparatively best model will not be the ‘best’ model because it is always impossible or at least difficult to justify the runtime to find the ‘best’ in a large pool of possible results.

\section*{8 Testing an Inverted Model}

Throughout this article, it has been discussed that BERTopic risks being a ‘black box’ (Wang et al., 2024). This is a common issue in digital humanities, where users may not understand what is happening within the package (Cigliano, Fallucchi and Gerardi, 2024; Wang et al., 2024). This makes it difficult for the average user to fine-tune their models or understand the cause of poor quality outputs. Just as they may be able to customise parameters, the wrapper is not modular. This means that fine-tuning through trial and error will be time-consuming. On one hand, this is not an issue with BERTopic but with how we have chosen to use it. The advantage of this model is that it provides summarising of large amounts of textual data reasonably quickly (Didehkhani, 2024), compared to manual reading. In tests for this project, BERTopic has run around 30 minutes with no pre-computed embeddings and around 2 hours for 30 runs with pre-computed embeddings. However, this process is more computationally demanding than LDA (Kumar, Karamchandani and Singh, 2024). It also uses HDBScan, a reliable clustering package, as demonstrated by the resilience to stress testing. The issue has been that, through changing parameters randomly, it has not been possible to increase the ngram frequency values. It has not been possible to change the models such that they have more high frequency ngrams.

Fundamentally, the models themselves, which we mean the ngram outputs and topic names, have remained reasonably consistent. While this relationship has been explored through ngram values and NUV scores, we have a further test using a fuzzy topic name test. This is designed to compare the topic names of the models to a list of precomputed names using WER to test how similar the topic names are. It outputs a value which is created through first finding how many previous topic names each of the top 20 names is meaningfully similar to. Then, cumulating the values.

The results found Spearman's Rho = -0.06, $p = 0.7137$. Therefore, there is no statistically significant relationship between topic name values (stability score) and sentence coverage. Once again, this suggests BERTopic is resilient to stress testing and provides consistent outputs. On the other hand, it suggests a greater issue that the BERTopic outputs are not changing when the clustering changes. This suggests the model is not representing shifts in topic clusters accurately. Since topic models are essential researcher constructs, the researchers set the rules of how the clusters are interpreted and what is the topic output, this is not surprising. However, it suggests an issue where attempts to rationalise unwieldy clusters are too discriminatory, or not sensitive enough to variation within clusters. Fundamentally, this becomes more a philosophical issue as to what constitutes the topic and how best might a cluster be interpreted. Ultimately, these clusters have been artificially manipulated to become viable inputs for HDBScan through UMAP. HDBScan is not designed for linguistic data, but any data. So therefore, using an unsupervised approach there are genuine risks of inappropriate usage of tools and improper metrics utilised to understand how effectively the ngram extraction is performing the desired task.

For an empirical comparison, what if we were to invert the process of BERTopic, beginning with high frequency ngrams and creating supervised clusters based on anchor sentences. Instead of relying on HDBScan for clustering, we can compare the outputs from this approach to a BERTopic output to see how similar the results are. This could be useful for exploring what effect an approach which focuses on high value ngrams would be. As previously suggested, there is reason to be concerned that using this metric will lead to better outcomes. Some high-frequency ngrams will not be thematically useful for researchers. On the other hand, if BERTopic is providing reasonably stable labels, does this suggest these corpora may be suited to a deterministic approach? One finding so far has been resilience to researcher interferences, suggesting that using unsupervised models does not lead to a plurality of outcomes.

With the large dataset, the NUV was 0.67, under the BERTopic mean of 0.81. In fact, this is below the minimum value for BERTopic computed.

By using an inverse model, it may be possible to explore the issues within topic models without the ‘blackbox’ issue (Wang et al., 2024). In this new evaluation, we shall compare the effect of changing cluster cohesion on the pre-established metrics. This model performs semi-supervised clustering by finding representative sentences of high frequency ngrams and uses cosine similarity to use them as anchors for clusters. Through this approach it is possible to test how changing the threshold from 0.3 to 1 with an increment of 0.01. This should provide 70 values, from weak, large clusters at 0.3, to small clusters at 1. Before considering the results, it is worth recognising first that this model has a limitation of using sentences as representative of topics. To ensure each sentence is valid, sentences were chosen based on not including high value unigrams and only containing one high-value ngram. This should find unique sentences that are semantically unique, but not so semantically unique that there will be no similar sentences.

The goal of this approach is not to create a perfect model, but to use a method that is dissimilar to unsupervised clustering. At the same time, this model will provide the same outputs as BERTopic, providing the opportunity for exploration of how far this model faces the same trade-off as BERTopic and what this might suggest about the limitations of topic modelling more broadly.

\section*{9 Conclusions}

Despite many of the issues discussed in this article, it should not be overlooked that BERTopic is a strong topic model. It may be slow in clustering, non-deterministic and suffering from trade-offs from optimisation. On the other hand, it considers semantic information, effective at creating unique topics and able to set an appropriate number of topics from a corpus instead of having a preset cluster amount like K means. Moreover, it is customisable, modular and able to visualise the results effectively. The primary issue is that it is in danger of becoming a ‘blackbox’ where a researcher or any user would use BERTopic and assume any output is optimal (Benz et al., 2025; Wang et al., 2024).

There appears to be variation along the different metrics explored between better and worse models in certain areas. Equally, the choice of deciding which metrics to optimise leads to trade-offs that might ask greater questions of what topic models are for. Whether topic models exist to explain the most possible data or whether they exist to create the most coherent topic appear to be contradictory aims within this model. It will depend on one’s interpretation of explanatory power within a topic model. This is an issue because the number of sentences within the -1 may not be a useful metric. This is because topics above the count of 20 may be excluded by the user or may be of poor quality. Simplistically, the question is whether we are evaluating the topic model as a whole or the top 20 topics. If we are evaluating with the use cases in mind, it would seem strange to be evaluating beyond the 20 mark. This is why this article has focused on evaluations of the top 20. Equally, because of how the sentence counts can decrease already by the 20th topic, these top 20 are liable to include the majority of topics. With this in context, the damage of not adding topics past the 20th to the -1 number of sentences may not be such an issue. On the other hand, the issue is not entirely quantitative. It could be possible that the -1 is a reflection that every corpus will have low-value sentences. The issue with this argument is that it would be impossible to create a standard for this. Semantic outliers can easily be identified but these are not necessarily low value. What we do not know from the -1 category is how appropriate that count is. It is feasible that it could be that each corpus should have a number of sentences in this category.


\begin{thebibliography}{10}

\bibitem{An2023}
An, Y., Oh, H., \& Lee, J.
\newblock Marketing insights from reviews using topic modeling with BERTopic and deep clustering network.
\newblock {\em Applied Sciences}, 13(16):9443, 2023.
\newblock \href{https://doi.org/10.3390/app13169443}{\texttt{DOI: 10.3390/app13169443}}.
\newblock [Online]. Accessed: 18 July 2025.

\bibitem{Cigliano2024}
Cigliano, A., Fallucchi, F., \& Gerardi, M.
\newblock The impact of digital analysis and large language models in digital humanity.
\newblock In {\em ICYRIME 2024: 9th International Conference of Yearly Reports on Informatics, Mathematics, and Engineering}, volume 3869 of {\em CEUR Workshop Proceedings}, pages 1--6, 2024.
\newblock \href{https://ceur-ws.org/Vol-3869/p01.pdf}{\texttt{URL}}.
\newblock [Online]. Accessed: 18 July 2025.

\bibitem{Didehkhani2024}
Didehkhani, Z.
\newblock Analyzing Persian Twitter sentiments on the Arbaeen walk: A comparative study of LDA and BERTopic with the Arbaeen tweets dataset.
\newblock 2024.
\newblock \href{https://mlkd.aut.ac.ir/proceedings/2024/paper/2A.2.pdf}{\texttt{URL}}.
\newblock [Online]. Accessed: 18 July 2025.

\bibitem{Egger2022}
Egger, R. \& Yu, J.
\newblock A topic modeling comparison between LDA, NMF, Top2Vec, and BERTopic to demystify Twitter posts.
\newblock {\em Frontiers in Sociology}, 7:886498, 2022.
\newblock \href{https://doi.org/10.3389/fsoc.2022.886498}{\texttt{DOI: 10.3389/fsoc.2022.886498}}.
\newblock [Online]. Accessed: 18 July 2025.

\bibitem{Gan2024}
Gan, L., Lu, H., \& Cai, J.
\newblock Experimental comparison of three topic modeling methods with LDA, Top2Vec and BERTopic.
\newblock In H. Lu \& J. Cai, editors, {\em Artificial Intelligence and Robotics}, volume 1998 of {\em Communications in Computer and Information Science}, pages 376--391. Springer, Singapore, 2024.
\newblock \href{https://doi.org/10.1007/978-981-99-9109-9_37}{\texttt{DOI: 10.1007/978-981-99-9109-9\_37}}.
\newblock [Online]. Accessed: 18 July 2025.

\bibitem{Grootendorst2022}
Grootendorst, M.
\newblock BERTopic: Neural topic modeling with a class-based TF-IDF procedure.
\newblock 2022.
\newblock \href{https://doi.org/10.48550/arXiv.2203.05794}{\texttt{arXiv:2203.05794}}.
\newblock [Online]. Accessed: 18 July 2025.

\bibitem{HosseinyMarani2023}
Hosseiny Marani, A. \& Baumer, E.~P.~S.
\newblock A review of stability in topic modeling: Metrics for assessing and techniques for improving stability.
\newblock {\em ACM Computing Surveys}, 56(5):108:1--108:32, 2023.
\newblock \href{https://doi.org/10.1145/3623269}{\texttt{DOI: 10.1145/3623269}}.
\newblock [Online]. Accessed: 18 July 2025.

\bibitem{Khodeir2025}
Khodeir, N. \& Elghannam, F.
\newblock Efficient topic identification for urgent MOOC Forum posts using BERTopic and traditional topic modeling techniques.
\newblock {\em Education and Information Technologies}, 30(5):5501--5527, 2025.
\newblock \href{https://doi.org/10.1007/s10639-024-13003-4}{\texttt{DOI: 10.1007/s10639-024-13003-4}}.
\newblock [Online]. Accessed: 18 July 2025.

\bibitem{Kumar2024}
Kumar, A., Karamchandani, A., \& Singh, S.
\newblock Topic modeling of neuropsychiatric diseases related to gut microbiota and gut brain axis using artificial intelligence based BERTopic model on PubMed abstracts.
\newblock {\em Neuroscience Informatics}, 4(4):100175, 2024.
\newblock \href{https://doi.org/10.1016/j.neuri.2024.100175}{\texttt{DOI: 10.1016/j.neuri.2024.100175}}.
\newblock [Online]. Accessed: 18 July 2025.

\bibitem{Li2023}
Li, H. et~al.
\newblock Research on a data mining algorithm based on BERTopic for medication rules in Traditional Chinese Medicine prescriptions.
\newblock {\em Medicine Advances}, 1(4):353--360, 2023.
\newblock \href{https://doi.org/10.1002/med4.39}{\texttt{DOI: 10.1002/med4.39}}.
\newblock [Online]. Accessed: 18 July 2025.

\bibitem{Li2025}
Li, X. et~al.
\newblock Evaluation of unsupervised static topic models' emergence detection ability.
\newblock {\em PeerJ Computer Science}, 11:e2875, 2025.
\newblock \href{https://doi.org/10.7717/peerj-cs.2875}{\texttt{DOI: 10.7717/peerj-cs.2875}}.
\newblock [Online]. Accessed: 18 July 2025.

\bibitem{Lin2024}
Lin, Q. et~al.
\newblock Research on topic mining and evolution trends of functional agriculture based on the BERTopic model.
\newblock {\em Agriculture}, 14(10):1691, 2024.
\newblock \href{https://doi.org/10.3390/agriculture14101691}{\texttt{DOI: 10.3390/agriculture14101691}}.
\newblock [Online]. Accessed: 18 July 2025.

\bibitem{MadridGarcia2024}
Madrid-García, A. et~al.
\newblock Mapping two decades of research in rheumatology-specific journals: a topic modeling analysis with BERTopic.
\newblock {\em Therapeutic Advances in Musculoskeletal Disease}, 16:1759720X241308037, 2024.
\newblock \href{https://doi.org/10.1177/1759720X241308037}{\texttt{DOI: 10.1177/1759720X241308037}}.
\newblock [Online]. Accessed: 18 July 2025.

\bibitem{Medvecki2024}
Medvecki, D. et~al.
\newblock Multilingual transformer and BERTopic for short text topic modeling: The case of Serbian.
\newblock In {\em Disruptive Information Technologies for a Smart Society}, volume 872 of {\em Lecture Notes in Networks and Systems}, pages 161--173. Springer, 2024.
\newblock \href{https://doi.org/10.1007/978-3-031-50755-7_16}{\texttt{DOI: 10.1007/978-3-031-50755-7\_16}}.
\newblock [Online]. Accessed: 18 July 2025.

\bibitem{Ranaweera2025}
Ranaweera, N. et~al.
\newblock BERTDetect: A neural topic modelling approach for Android malware detection.
\newblock In {\em Companion Proceedings of the ACM on Web Conference 2025}, pages 1802--1810, 2025.
\newblock \href{https://doi.org/10.1145/3701716.3717501}{\texttt{DOI: 10.1145/3701716.3717501}}.
\newblock [Online]. Accessed: 18 July 2025.

\bibitem{Wang2024}
Wang, X. et~al.
\newblock Digital deduction theatre: An experimental methodological framework for the digital intelligence revitalisation of cultural heritage.
\newblock In {\em Intelligent Computing for Cultural Heritage}, pages 203--220, 2024.
\newblock \href{https://library.oapen.org/bitstream/handle/20.500.12657/92133/9781040113264.pdf?sequence=1#page=232}{\texttt{URL}}.
\newblock [Online]. Accessed: 18 July 2025.

\bibitem{Zhang2024}
Zhang, N. \& Wang, J.
\newblock Topic analysis of digital preservation based on BERTopic.
\newblock In {\em iPRES 2024}, 2024.
\newblock \href{https://doi.org/10.21428/5676bf2d.6f3ce886}{\texttt{DOI: 10.21428/5676bf2d.6f3ce886}}.
\newblock [Online]. Accessed: 18 July 2025.

\end{thebibliography}
\end{document}